\documentclass[aps,pra,twocolumn,superscriptaddress,showpacs,amsmath,amssymb,notitlepage]{revtex4-1}

\usepackage{amsfonts}
\usepackage{amssymb}
\usepackage{graphicx}
\usepackage{color}

\usepackage{bm}
\usepackage{braket}

\usepackage{hyperref}

\usepackage[normalem]{ulem}

\begin{document}
\title{Bloch oscillations in the spin-1/2 XXZ chain}
\author{Yankang Liu}
\affiliation{Department of Applied Physics, University of Tokyo, Tokyo 113-8656, Japan}
\author{Yohei Fuji}
\affiliation{Department of Applied Physics, University of Tokyo, Tokyo 113-8656, Japan}
\author{Haruki Watanabe}\email{hwatanabe@g.ecc.u-tokyo.ac.jp}
\affiliation{Department of Applied Physics, University of Tokyo, Tokyo 113-8656, Japan}

\begin{abstract}
Under a perfect periodic potential, the electric current density induced by a constant electric field may exhibit nontrivial oscillations, so-called Bloch oscillations, with an amplitude that remains nonzero in the large system size limit. Such oscillations have been well studied for nearly noninteracting particles and observed in experiments. In this work, we revisit  Bloch oscillations in strongly interacting systems.  
By analyzing the spin-1/2 XXZ chain, which can be mapped to a model of spinless electrons, we demonstrate that the current density at special values of the anisotropy parameter $\Delta=-\cos(\pi/p)$ ($p=3,4,5,\cdots$) in the ferromagnetic gapless regime behaves qualitatively the same as in the noninteracting case ($\Delta=0$) even in the weak electric field limit. When $\Delta$ deviates from these values, the amplitude of the oscillation under a weak electric field is suppressed by a factor of the system size.   We estimate the strength of the electric field required to observe such a behavior using the Landau--Zener formula.
\end{abstract}

\maketitle

\section{Introduction}
\begin{figure}[t]
\begin{center}
\includegraphics[width=0.99\columnwidth]{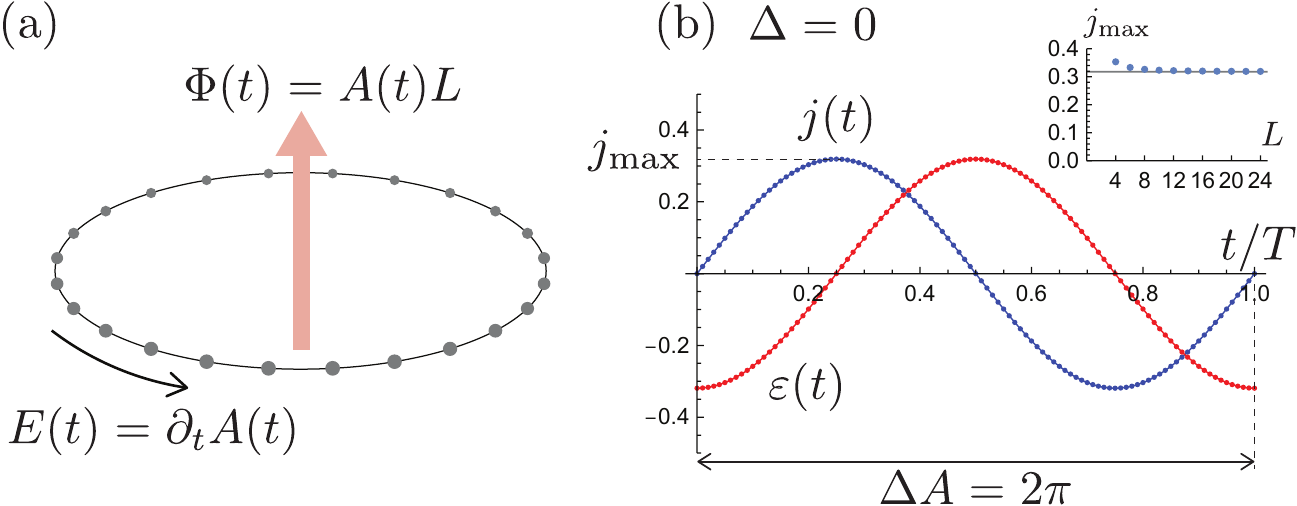}
\caption{
(a) One-dimensional ring pierced by a flux $\Phi(t)=A(t)L$, which induces an electric field $E(t)=\partial_tA(t)$. 
(b) Oscillations of the energy density $\varepsilon(t)$ in \eqref{defEt} and the current density $j(t)$ in \eqref{defjt} for the model \eqref{Ht} with $A=A(t)=2\pi t/T$ and $L=24$ in the noninteracting limit $\Delta=0$.  We set $T=1$ in this plot but results are almost identical up to $T=10^6$, the largest period we computed.
Both $j(t)$ and $\varepsilon(t)$ show a single oscillation over the period $T$ where $A$ increases by $2\pi$.
Solid curves represent the analytic expressions $\varepsilon(t)=-L^{-1}\cos A(t)/\sin(\pi/L)$ and $j(t)=L^{-1}\sin A(t)/\sin(\pi/L)$.
The inset shows the amplitude $j_{\text{max}}$ as a function of $L$, suggesting that it is of $O(1)$ in the large $L$ limit.
}
\label{fig1}
\end{center}
\end{figure}

Imagine a system of electrons in a one-dimensional ring [Fig~\ref{fig1}(a)]. Suppose that a static and uniform electric field $E$ is turned on at time $t=0$.  When $E$ is sufficiently weak, 
the response of the electric current density $j(t)$ is dominated by the contribution from the Drude weight $\mathcal{D}_1$ of the linear response theory~\cite{Resta_2018}. In free space, electrons keep accelerated and the induced electric current density increases linearly as a function of time, i.e., $j(t)\simeq \mathcal{D}_1 Et$. In the presence of a disorder-free lattice with the lattice constant $a$, on the other hand, noninteracting electrons get accelerated initially but then start to move backward [Fig.~\ref{fig1}(b)], forming an oscillatory motion with the frequency $aE/(2\pi)$ (we set $e=\hbar=1$  throughout this work). Such an oscillation is known as a Bloch oscillation and has been observed in semiconductor superlattices~\cite{PhysRevB.46.7252,LEO1992943,PhysRevLett.70.3319,PhysRevB.51.17275}. The oscillation can be understood as a result of nonlinear responses of the electric current to higher powers of $E$~\cite{PhysRevB.102.165137}.

The adiabatic transport property is known to be strongly affected by interactions and disorders~\cite{KaneFisher-PRL1992,KaneFisher-PRB1992}.
For interacting systems, recent studies revealed that the Drude weight $\mathcal{D}_N$ $(N\geq2)$ of nonlinear responses tends to diverge in the limit of large system size~\cite{PhysRevB.102.165137,PhysRevB.103.L201120,Fava}.  Similar divergence has been observed in
noninteracting systems in the presence of a localized impurity~\cite{TakasanOshikawaWatanabe}.  These results imply that the Bloch oscillations in noninteracting and imperfection-free systems may completely disappear when interactions or disorders are added. This is indeed the case for a tight-binding model with a single defect~\cite{TakasanOshikawaWatanabe}.

In this work, we study the current response of disorder-free interacting electrons in the limit of weak electric field.
By analyzing the spin-1/2 XXZ chain, which can be mapped to a model of spinless electrons interacting with each other between neighboring sites, we find that interacting systems can exhibit Bloch oscillations for fine-tuned values of the interaction $\Delta=-\cos(\pi/p)$  ($p=3,4,5,\cdots$). When $\Delta$ is not exactly at these special values, the interaction induces level repulsions to the many-body energy levels as a function of the gauge field $A$. As a result, the behavior of $j(t)$ becomes qualitatively different --- it still oscillates as a function of $t$ but with an amplitude inversely proportional to the system size. At the same time, the frequency becomes $LE/(4\pi)$ and is proportional to the system size.  Such behaviors can be seen only when the electric field $E$ is sufficiently weak, and we give an estimate of the required strength of $E$ using the Landau--Zener formula~\cite{landau1932theory,doi:10.1098/rspa.1932.0165,doi:10.1021/jp040627u}.

\section{The XXZ model}
\subsection{Definitions}
In this work, we discuss the spin-1/2 XXZ chain
\begin{equation}
\hat{H}(A)=J\sum_{i=1}^L\left(\frac{1}{2}\hat{s}_{i+1}^+e^{-iA}\hat{s}_i^-+\text{h.c.}+\Delta\hat{s}_{i+1}^z\hat{s}_i^z\right),
\label{Ht}
\end{equation}
where 
$\hat{s}_i^\alpha$ ($\alpha=x,y,z$) is the spin-1/2 operator defined on the site $i$ and $\hat{s}_{i}^\pm=\hat{s}_{i}^x\pm i\hat{s}_{i}^y$, 
$L$ is the system size,
$J$ is the antiferromagnetic coupling constant, 
$\Delta$ is the anisotropy parameter, and 
$A$ is the vector potential (with the sign convention opposite to the conventional one) associated to the spin rotation symmetry about $z$ axis. 
The lattice constant $a$ is set to $1$.
We impose the periodic boundary condition and identify $\hat{s}^{\alpha}_{L+1} = \hat{s}^{\alpha}_{1}$.  The conserved current density of this model is defined by
\begin{equation}
\hat{j}(A)=\frac{1}{L}\frac{d\hat{H}(A)}{dA}.
\label{jt}
\end{equation}

The spin chain~\eqref{Ht} can be mapped to a model of spinless electrons
\begin{align}
\hat{H}(A)&=\sum_{i=1}^L\left(t_0\hat{c}_{i+1}^\dagger e^{-iA}\hat{c}_i +\text{h.c.}\right)\notag\\
&\quad+\sum_{i=1}^LU\Big(\hat{n}_{i+1}-\frac{1}{2}\Big)\Big(\hat{n}_i-\frac{1}{2}\Big)
\end{align}
by the Jordan--Wigner transformation~\cite{doi:10.1142/S0217979212440092}, where $t_0=J/2$ and $U=J\Delta$. The spin current density \eqref{jt} can thus be regarded as the electric current density of spinless electrons.  In the following, we set $J=1$ and focus on the gapless regime $-1<\Delta<1$.  

\subsection{Periodicity of the adiabatic ground state}
Let us examine the periodicity of the many-body energy levels of $\hat{H}(A)$ as a function of $A$. Obviously we have $\hat{H}(2\pi)=\hat{H}(0)$, but $\Delta A=2\pi$ is not the smallest period of this model in the following sense.  The gauge field $A$ is related to the flux $\Phi=AL$ piercing the hole of the 1D ring [Fig.~\ref{fig1}(a)], and $A=0$ and $A=2\pi/L$ give the physically equivalent flux. In fact, they are connected by a unitary transformation, called the large gauge transformation~\cite{Oshikawa2000}. Therefore, the entire spectrum of $\hat{H}(A)$ has a period $\Delta  A=2\pi/L$. This is the case regardless of the specific choice of $\Delta$. 

\begin{figure}[t]
\begin{center}
\includegraphics[width=0.99\columnwidth]{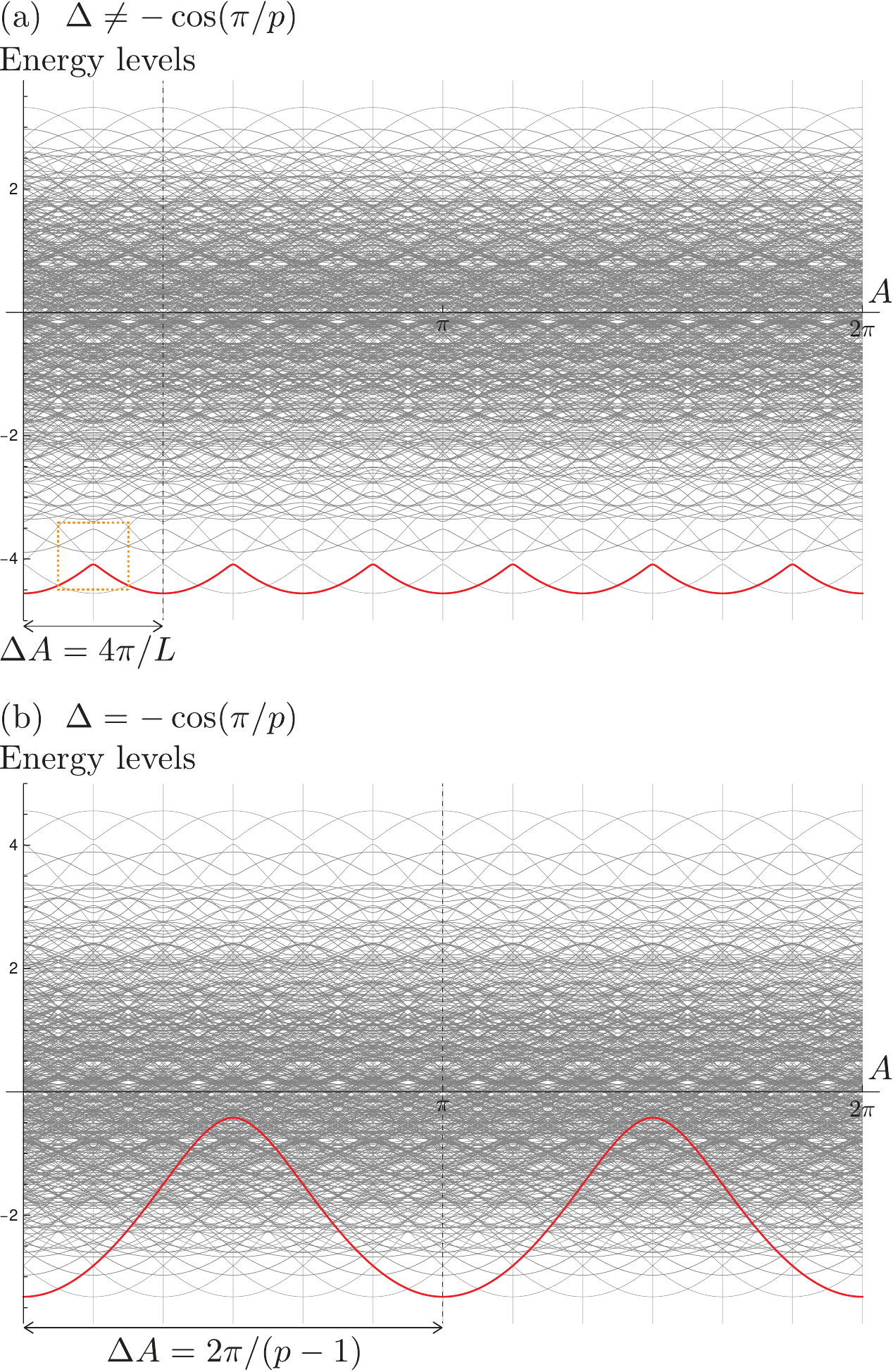}
\caption{The many-body energy levels of $\hat{H}(A)$, obtained by the exact diagonalization, for $L=12$ as a function of the gauge field $A$ in the case of (a) a generic value $\Delta=+0.5$ and (b) a special value $\Delta=-\cos(\pi/3)$ ($p=3$). 
Only the states in the $S_z=0$ sector are shown. The adiabatic ground state is highlighted with red color.
Because $\Delta\leftrightarrow -\Delta$ corresponds to $\hat{H}(A)\leftrightarrow -\hat{H}(A)$, (b) can be seen as the inverted version of (a) in this example.
}
\label{fig2}
\end{center}
\end{figure}

In contrast, the ``adiabatic ground state" has a longer period that depends on $\Delta$~\cite{ALCARAZ1988280}.
Let $|\Psi\rangle$ be the ground state at $A=0$. Among the stroboscopic eigenstates of $\hat{H}(A)$, we keep track of the state that is smoothly connected to $|\Psi\rangle$ as a function of $A$. This is what we call the adiabatic ground state $|\Psi_{\text{AGS}}(A)\rangle$. See red curves in Fig.~\ref{fig2}  for illustration. In the following, we denote by $\varepsilon_{\text{AGS}}(A)$ the energy density of the adiabatic ground state.

The flux-threading argument for the Lieb-Schultz-Mattis theorem predicts a level crossing between two levels with distinct momentum eigenvalues $0$ and $\pi$
for translation and time-reversal invariant spin-$1/2$ chains~\cite{Oshikawa2000}.  Hence, the smallest possible values of the period of the adiabatic ground state is $\Delta  A=4\pi/L$. This is indeed the case for generic $\Delta$~\cite{PhysRevLett.65.1833,doi:10.1143/JPSJ.65.2824} [Fig.~\ref{fig2}(a)]. In particular, the period $\Delta  A$ vanishes in the large $L$ limit and the energy density $\varepsilon_{\text{AGS}}(A)$ becomes a constant $\varepsilon_{\text{AGS}}(0)$ independent of $A$.

\begin{figure}[t]
\begin{center}
\includegraphics[width=0.99\columnwidth]{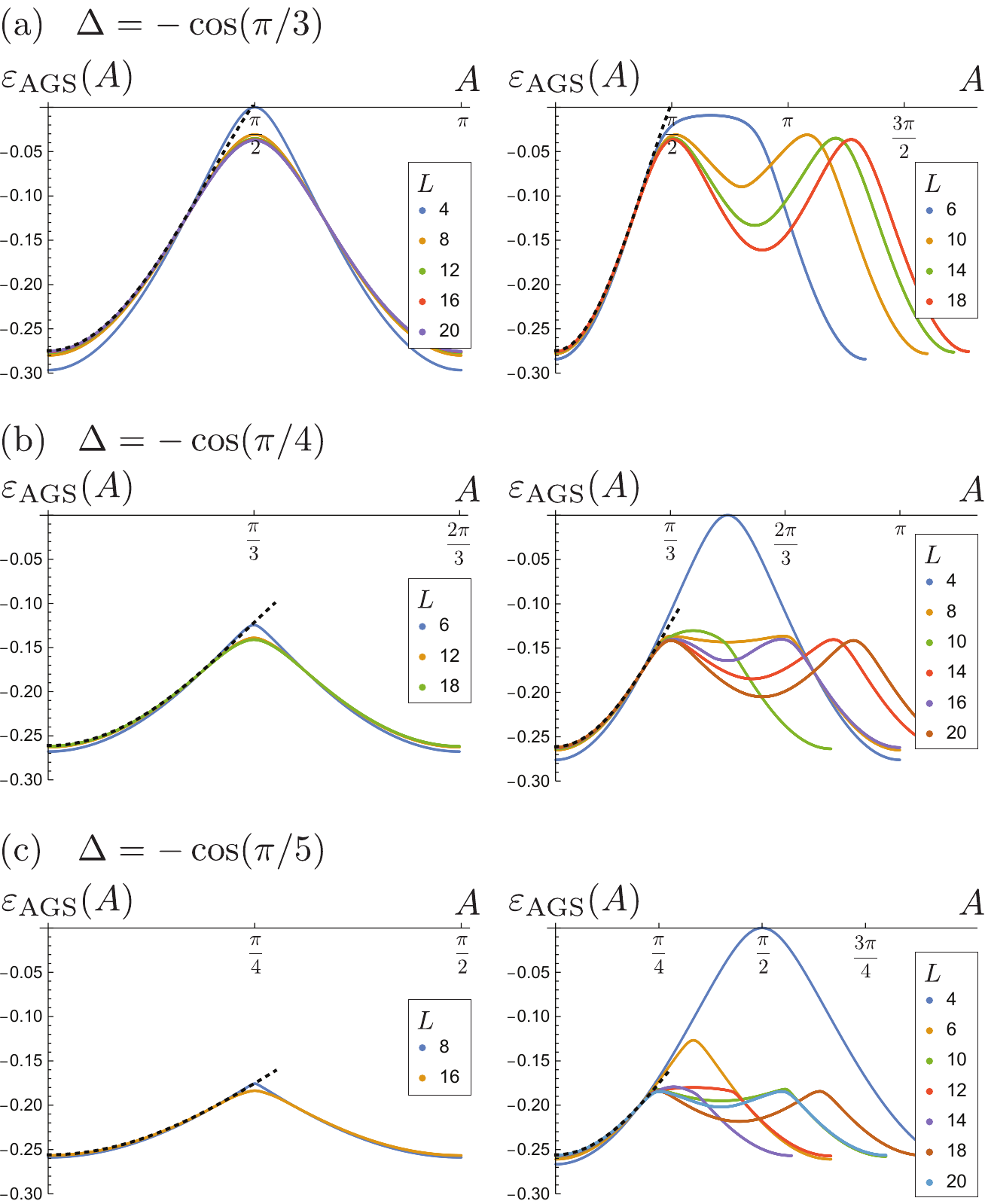}
\caption{The energy density of the adiabatic ground state $\varepsilon_{\text{AGS}}(A)$ as a function of $A$ for (a) $\Delta=-\cos(\pi/3)=-0.5$,  (b) $\Delta=-\cos(\pi/4)=-0.707107$, and  (c) $\Delta=-\cos(\pi/5)=-0.809017$. The dotted curve is the analytic expression for $L\rightarrow\infty$ up to the $A^6$ order in \eqref{exp6}.
}
\label{fig3}
\end{center}
\end{figure}

Exceptions occur at special values of $\Delta$, i.e., $\Delta=-\cos(\pi/p)$ ($p=2,3,\cdots$), at which the adiabatic ground state has a much longer period~\cite{PhysRevB.46.14583}, as shown in Fig.~\ref{fig2}(b) for $p=3$. From numerical investigations, we identify the period $\Delta A=2\pi/(p-1)$ for the sequence $L=2(p-1)\ell$ ($\ell\in\mathbb{N}$)~\footnote{There is a factor of four error in Ref.~\onlinecite{PhysRevB.46.14583}, which stated on the page 14590 that the period of $\Phi=AL$ is $8\pi L/(p-1)$ for $\Delta=\cos\mu=-\cos(\pi/p)$ with $p=\pi/(\pi-\mu)\in\mathbb{N}$ in our notation and the sign convention of $\Delta$. Our formula $\Delta A=2\pi/(p-1)$ gives the correct period $2\pi$ in the noninteracting case ($p=2$).}.  The system size dependence is examined in more detail by the exact diagonalization in Fig.~\ref{fig3}.  
For example, in the case of $p=3$, the period is $\Delta A=\pi$ for $L=4\ell$ and $\Delta A=2\pi(L-2)/L$ for $L=4\ell+2$.  The actual period converges to $\Delta A=\pi$ in the large $L$ limit for the latter case as well [see Fig.~\ref{fig3}(a)]. As shown in Fig.~\ref{fig3}(b) and (c), similar behaviors are also observed for the $p=4$ and $p=5$ cases.
Most importantly, both the period $\Delta  A$ and the amplitude of the oscillation of $\varepsilon_{\text{AGS}}(A)$ do not vanish in the large $L$ limit.
This difference results in the qualitatively different behaviors of the induced electric current in response to an infinitesimal electric field, as we discuss in the next section.

In Fig~\ref{fig3}, the dotted curve represents 
\begin{equation}
\varepsilon_{\text{AGS}}(A)\simeq\bar{\varepsilon}_{\text{GS}}+\frac{\bar{\mathcal{D}}_{1}}{2!} A^{2}+\frac{\bar{\mathcal{D}}_{3}}{4!} A^{4}+\frac{\bar{\mathcal{D}}_{5}}{6!} A^{6},\label{exp6}
\end{equation}
where $\bar{\varepsilon}_{\text{GS}}$ is the energy density and $\bar{\mathcal{D}}_{n}$ is the $n$-th order Drude weights in the large $L$ limit.
The analytic expressions of $\bar{\varepsilon}_{\text{GS}}$ and $\bar{\mathcal{D}}_{1}$ have long been known~\cite{Hamer_1987,PhysRevLett.65.1833}.
The expressions for higher Drude weights, $\bar{\mathcal{D}}_{3}$ and $\bar{\mathcal{D}}_{5}$, have been obtained in recent studies~\cite{PhysRevB.102.165137,PhysRevB.103.L201120}.
For readers' convenience, we include these expressions in Appendix~\ref{sec:app}.
We see an excellent agreement for small $A$, supporting the numerical stability of our results. 
Assuming that $\varepsilon_{\text{AGS}}(A)$ converges to a smooth function of $A$ in the large $L$ limit, all the nonlinear Drude weights remain finite for the special values of $\Delta$~\cite{TakasanOshikawaWatanabe}.

Before closing this section, let us explain how we kept track of the adiabatic ground state $|\Psi_{\text{AGS}}(A)\rangle$ in our numerical investigation of $\varepsilon_{\text{AGS}}(A)$. First, we reduce the number of candidate states by using the spin rotation symmetry about $z$-axis and the translation symmetry, focusing on the $S_z=0$ sector and the momentum $P=0$ sector. Next, we discretize $A\in[0,2\pi]$ as $A^{(n)}\equiv n\,\delta A$ ($n=0,1,2,\cdots$). The increment $\delta A$ needs to be chosen sufficiently small, and in this work we set $\delta A=\pi/(160L)$. We start with the unique ground state of $\hat{H}(A)$ for $A=A^{(0)}=0$ and $A=A^{(1)}=\delta A$. Then,  for $n>1$, we proceed step by step by choosing as $|\Psi_{\text{AGS}}(A^{(n+1)})\rangle$ the state whose energy density is closest to the linear interpolation
\begin{equation}
\varepsilon_{\text{AGS}}(A^{(n)})+\delta A[\varepsilon_{\text{AGS}}(A^{(n)})-\varepsilon_{\text{AGS}}(A^{(n-1)})]
\end{equation}
among the eigenstates of $\hat{H}(A^{(n+1)})$ in the $S_z=P=0$ sector.
We confirmed the convergence by changing $\delta A$.

\section{Time evolution}
\subsection{Bloch oscillations in the weak field limit}
Now, following Ref.~\cite{PhysRevLett.90.236401}, let us introduce a time dependence to the Hamiltonian \eqref{Ht} by setting $A=A(t)=2\pi t/T$ for $t\geq0$ and $A=0$ for $t<0$. 
This time-dependent gauge field induces a static and uniform electric field $E=2\pi /T$ for $t\geq0$.
We assume that the system is in the ground state $|\Psi\rangle$ of $\hat{H}(0)$ for $t<0$.
The time evolution of the state for $t\geq0$ is described by the Schr\"odinger equation
\begin{equation}
i\frac{d}{dt}|\Psi(t)\rangle=\hat{H}(A(t))|\Psi(t)\rangle.
\label{TDSE}
\end{equation}
The energy density $\varepsilon(t)$ and the induced current density $j(t)$ and  at time $t>0$ are given by
\begin{align}
\varepsilon(t)&\equiv\frac{1}{L}\langle \Psi(t)| \hat{H}(A(t))|\Psi(t)\rangle,\label{defEt}\\
j(t)&\equiv\langle \Psi(t)| \hat{j}(A(t))|\Psi(t)\rangle=\frac{1}{L}\langle \Psi(t)| \frac{d\hat{H}(A)}{dA}|\Psi(t)\rangle\Big|_{A=A(t)}.\label{defjt}
\end{align}
When $T$ is sufficiently large for a given $L$ and $\Delta$, the adiabatic theorem~\cite{Kato,PhysRevLett.119.060201} gives~\cite{PhysRevB.102.165137}
\begin{align}
\varepsilon(t)&=\varepsilon_{\text{AGS}}(A(t)),\label{a1}\\
j(t)&=\frac{d\varepsilon_{\text{AGS}}(A)}{dA}\Big|_{A=A(t)}.\label{a2}
\end{align}
Namely, the energy density $\varepsilon(t)$ and the current density $j(t)$ of the time-dependent problem are fully characterized by the adiabatic ground-state energy density $\varepsilon_{\text{AGS}}(A)$ of the time-independent system.
This immediately implies the possibility of Bloch oscillations even under an infinitesimal electric field for the special values $\Delta=-\cos(\pi/p)$ ($p=2,3,\cdots$), where the period and the amplitude of the oscillations of $\varepsilon_{\text{AGS}}(A)$ remain nonzero in the large $L$ limit.

We demonstrate this prediction by numerically solving the Schr\"odinger equation using the fourth-order Runge--Kutta method. We mainly set $dt=0.01$ but sometimes use $dt=0.005$ if necessary. Our results are summarized in Fig.~\ref{fig4}. 
As anticipated, even for $T=10^6$, the amplitudes of the oscillations of the energy density $\varepsilon(t)$ and the current density $j(t)$ are of $O(1)$ for the special values of $\Delta=-\cos(\pi/p)$ in panels (a)--(c). In contrast, they are $O(L^{-1})$ for the generic case in (d). Over the time period $T$, the gauge field $A(t)$ increases by $2\pi$ and the number of peaks is $p-1$ for the special values  in panels (a)--(c)  and is of $O(L)$ for the generic case in (d). These are all consistent with the behavior of $\varepsilon_{\text{AGS}}(A)$ discussed in the previous section. 
Sharp drops of the electric current $j(t)$  around $t=2\pi(2j-1)/(LE)$ ($j=1,2,\dots$) seen in (d) underlie the divergence of nonlinear Drude weights for generic $\Delta$. 

\subsection{Landau--Zener formula}
When $\Delta$ deviates from the special values, the many-body energy spectrum opens a tiny gap $\Delta E$ between the adiabatic ground state and the next level at $A=2\pi/L$. See the orange dashed box in Fig.~\ref{fig2} (a). When $T$ is large enough, the applied electric field is sufficiently weak and the state $|\Psi(t)\rangle$ stays in the adiabatic ground state. In contrast, when $T$ gets smaller, non-adiabatic transition starts to occur.  This type of  tunneling has been studied for the Hubbard ring of spinful electrons in Refs.~\onlinecite{OkaAritaAoki2003,PhysRevB.86.075148}. Here we estimate the strength of the electric field $E$ required to avoid non-adiabatic transition in the spin-1/2 XXZ chain \ref{Ht} based on the Landau--Zener formula.

\begin{figure}[t]
\begin{center}
\includegraphics[width=0.99\columnwidth]{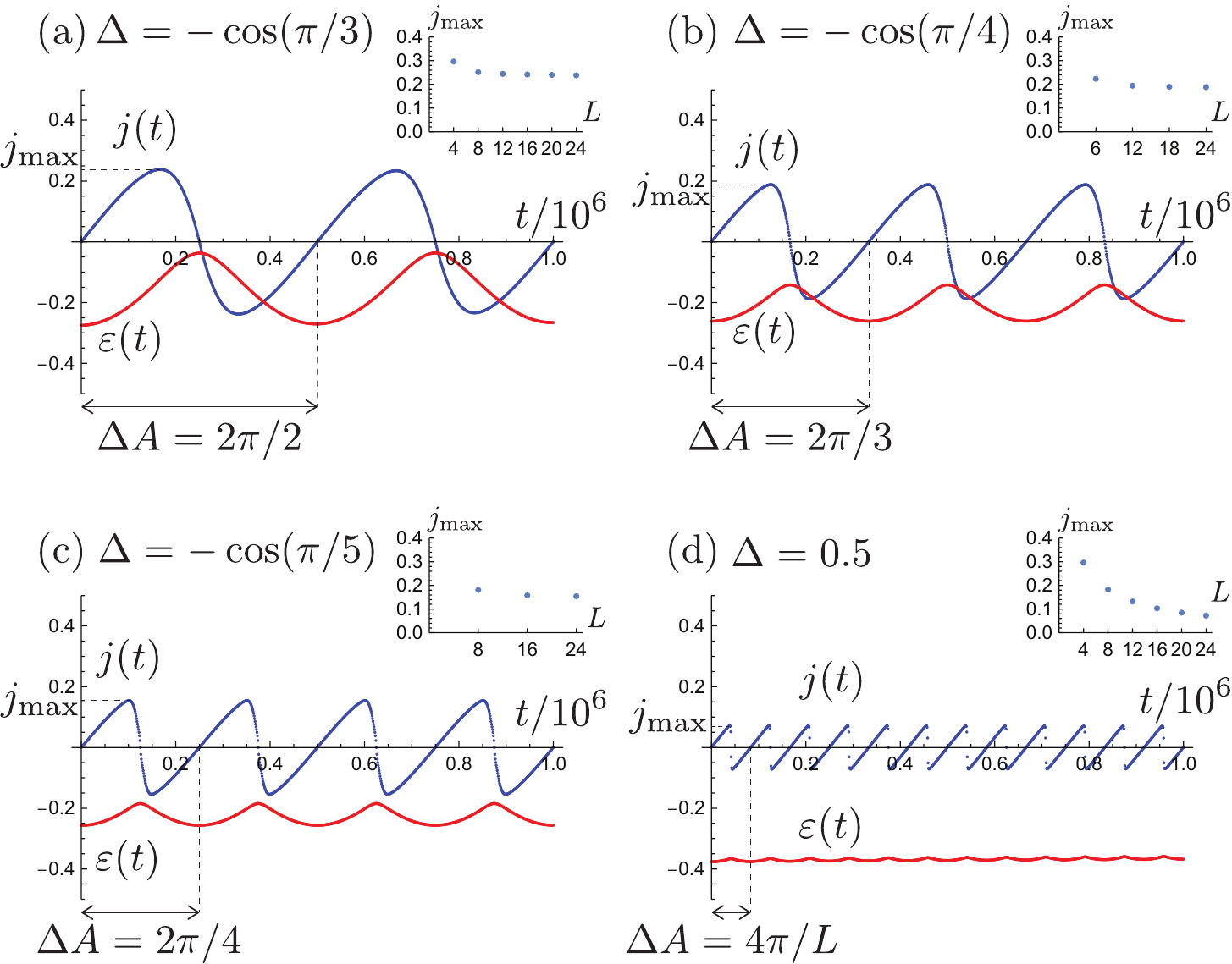}
\caption{Oscillations of the energy density $\varepsilon(t)$ in \eqref{defEt} and the current density $j(t)$ in \eqref{defjt} for 
(a) $\Delta=-\cos(\pi/3)=-0.5$, (b) $\Delta=-\cos(\pi/4)$, (c) $\Delta=-\cos(\pi/5)$, and (d) $\Delta=+0.5$.
We set $L=24$, $T=10^6$, and $A(t)=2\pi t/T$ in this calculation. The number of peaks is $p-1$ for (a)--(c) and $L/2$ for (d) as predicted. 
The inset shows the amplitude $j_{\text{max}}$ as a function of $L$, which remains nonzero in the large $L$ limit for (a)--(c), while it is inversely proportional to $L$ and vanishes  in the large $L$ limit for (d).
}
\label{fig4}
\end{center}
\end{figure}

\begin{figure}[t]
\begin{center}
\includegraphics[width=0.99\columnwidth]{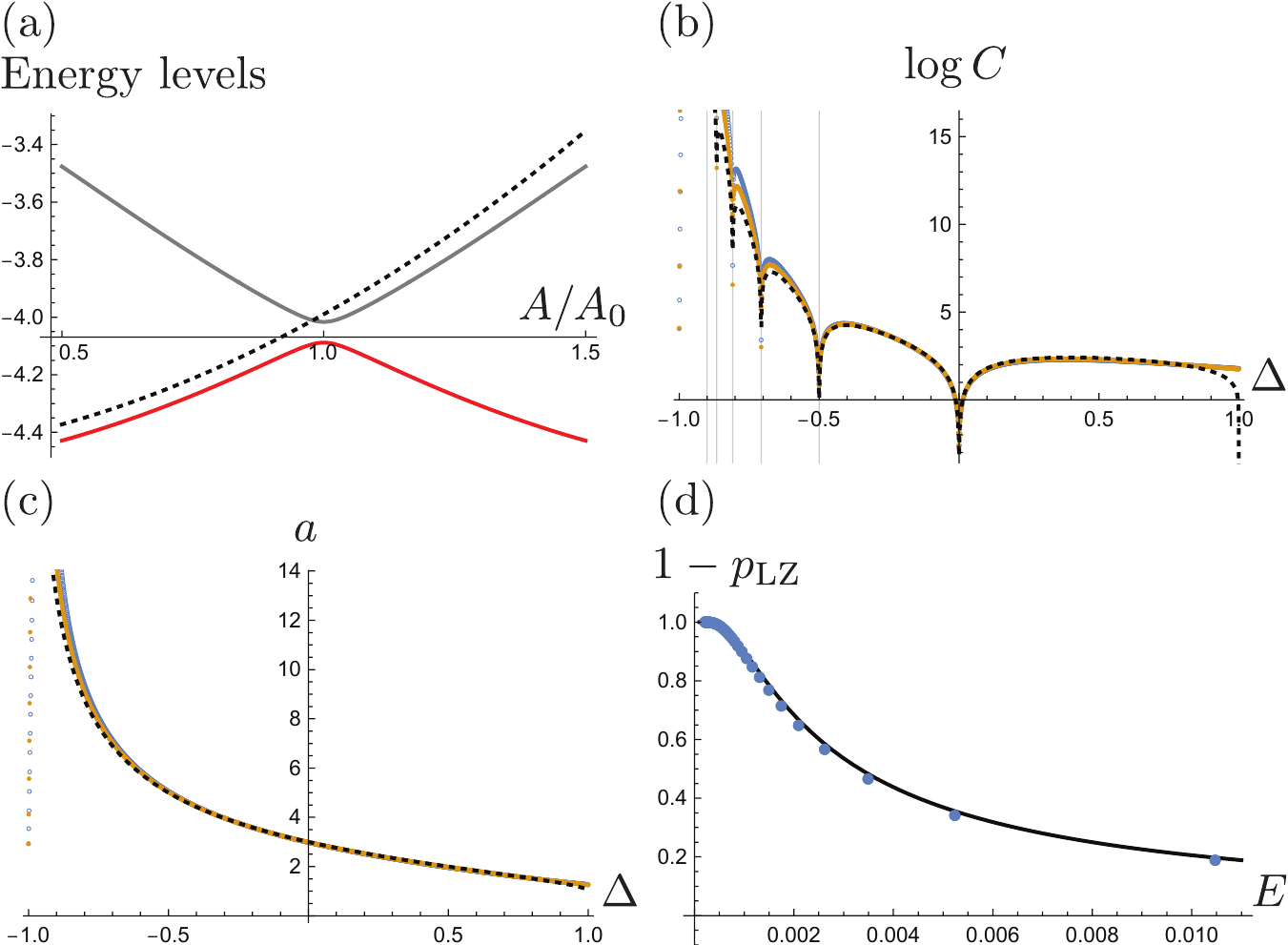}
\caption{(a) The zoom-up of the  orange dashed box in Fig.~\ref{fig2} (a), focusing on the adiabatic ground state and the next level near $A=A_0=2\pi/L$. The dashed curve represents $L\varepsilon_{\text{AGS}}(A)$ in the first line of \eqref{analitic2}.  
(b) The coefficient $C$ in \eqref{delta} obtained by fitting with numerical data.  The dotted curve represents the analytic expression in Appendix~\ref{sec:app}. The vertical lines represent $\Delta=-\cos(\pi/p)$ with $p=3,4,\dots,7$.
(c) The power $a$ in \eqref{delta} obtained by fitting.  The dotted curve represents $a=4K-1$. 
In panels (b) and (c), the blue open dots use data for $L=12, 14,\cdots,32$ and the yellow filled dots use data for $L=20, 22,\cdots,32$ only. 
(d) Numerical verification of the Landau--Zener formula for $\Delta=0.5$ and $L=12$. The blue dots represent $p_{\text{AGS}}(t)$ in \eqref{pAGS} at $t=2t_0$ and the solid curve represents $1-p_{\text{LZ}}$ given by \eqref{LZ}.
}
\label{fig6}
\end{center}
\end{figure}

Let us first assume $\Delta E=0$. Then the energy density of the adiabatic ground state around $A=A_0\equiv2\pi/L$ can be approximated, to the leading order of $L^{-1}$, by
\begin{align}
&\varepsilon_{\text{AGS}}(A)\simeq \bar{\varepsilon}_{\text{GS}}+\frac{1}{2}\bar{\mathcal{D}}_{1}A^2+O(A^4)\notag\\
&=\bar{\varepsilon}_{\text{GS}}+\frac{1}{2}\bar{\mathcal{D}}_{1}A_0^2+\bar{\mathcal{D}}_{1}A_0(A-A_0)+O((A-A_0)^2).\label{analitic2}
\end{align}
See Fig.~\ref{fig6}(a) for comparison with the exact diagonalization for $L=12$ and $\Delta=0.5$. The error in the approximation is a finite-size effect. Hence, when  $A(t)=Et$, the energy difference of the two levels that cross each other at $A=A_0$ goes as $\alpha|t-t_0|$ with $\alpha\equiv2\bar{\mathcal{D}}_1A_0LE=4\pi \bar{\mathcal{D}}_{1}E$ and $t_0\equiv A_0/E$.

When $\Delta E>0$, the Landau--Zener formula~\cite{landau1932theory,doi:10.1098/rspa.1932.0165,doi:10.1021/jp040627u} gives the probability of the non-adiabatic transition 
\begin{equation}
p_{\text{LZ}}\equiv\exp\left[-2\pi\frac{(\Delta E/2)^2}{\alpha}\right]=\exp\left[-\frac{(\Delta E)^2}{8\bar{\mathcal{D}}_{1}E}\right].\label{LZ}
\end{equation}
This formula was originally derived for two level systems and $\Delta E/2$ is identified with the amplitude of the off-diagonal components of the two level Hamiltonian.
Here, following Ref.~\onlinecite{OkaAritaAoki2003}, we apply it to the focused two levels in the many-body spectrum~\footnote{The Landau--Zener formula quoted in Eq.~(3) of Ref.~\onlinecite{OkaAritaAoki2003} missed a factor of $1/4$ in the exponent.}.

To make use of the formula \eqref{LZ}, we now study the system-size dependence of the finite-size splitting $\Delta E$. Here we assume the power-law form~\cite{PhysRevB.52.1656}
\begin{align}
\Delta E=CL^{-a}.\label{delta}
\end{align}
The leading contribution to $\Delta E$ comes from the least irrelevant perturbation due to Umklapp scattering~\cite{LUKYANOV1998533,doi:10.1143/JPSJ.65.2824}, which gives $a=4K-1=2\pi[\arccos(-\Delta)]^{-1}-1$, where $K$ is the Luttinger parameter~\cite{Giamarchi2004B}. We append the analytic expression for $C$ in Appendix \ref{sec:app}.  For example, we find $\Delta E\simeq 2\sqrt{3}\pi L^{-2}$ for $\Delta=0.5$. We confirm these analytic expressions by fitting with numerical data up to $L=32$. As shown in Fig~\ref{fig6} (b) and (c), we find a good agreement.

Finally, let us verify the Landau--Zener formula \eqref{LZ}. We compute the probability of adiabatic process in two ways: one by $1-p_{\text{LZ}}$ using the last expression of \eqref{LZ} and the other by the overlap
\begin{equation}
p_{\text{AGS}}(t)\equiv |\langle\Psi_{\text{AGS}}(A(t))|\Psi(t)\rangle|^2
\label{pAGS}
\end{equation}
at $t=2t_0$, by numerically solving the time-dependent Schr\"odinger equation \eqref{TDSE}.  For the gap $\Delta E\simeq C L^{-a}$, we use the analytic expressions of $C$ and $a$.  
As shown in Fig.~\ref{fig6}(d) for $\Delta=0.5$ and $L=12$,  they agree unexpectedly well.

These results suggest that, in order to avoid non-adiabatic transition, the electric field must satisfy
\begin{align}
E\ll \frac{(\Delta E)^2}{8\bar{\mathcal{D}}_{1}}=\frac{C^2}{8\bar{\mathcal{D}}_{1}}L^{-2a}.
\label{LZ2}
\end{align}
The right-hand side becomes small quickly as $L$ increases, suggesting that $E$ must be chosen very small. In other words, when $E=2\pi/T$, the time interval $T$ must be chosen quite long. 
Therefore, it is actually nearly impossible to achieve the adiabatic limit such as the one in Fig.~\ref{fig4}(d) in a thermodynamically large system.  

\begin{figure}[t]
\begin{center}
\includegraphics[width=0.99\columnwidth]{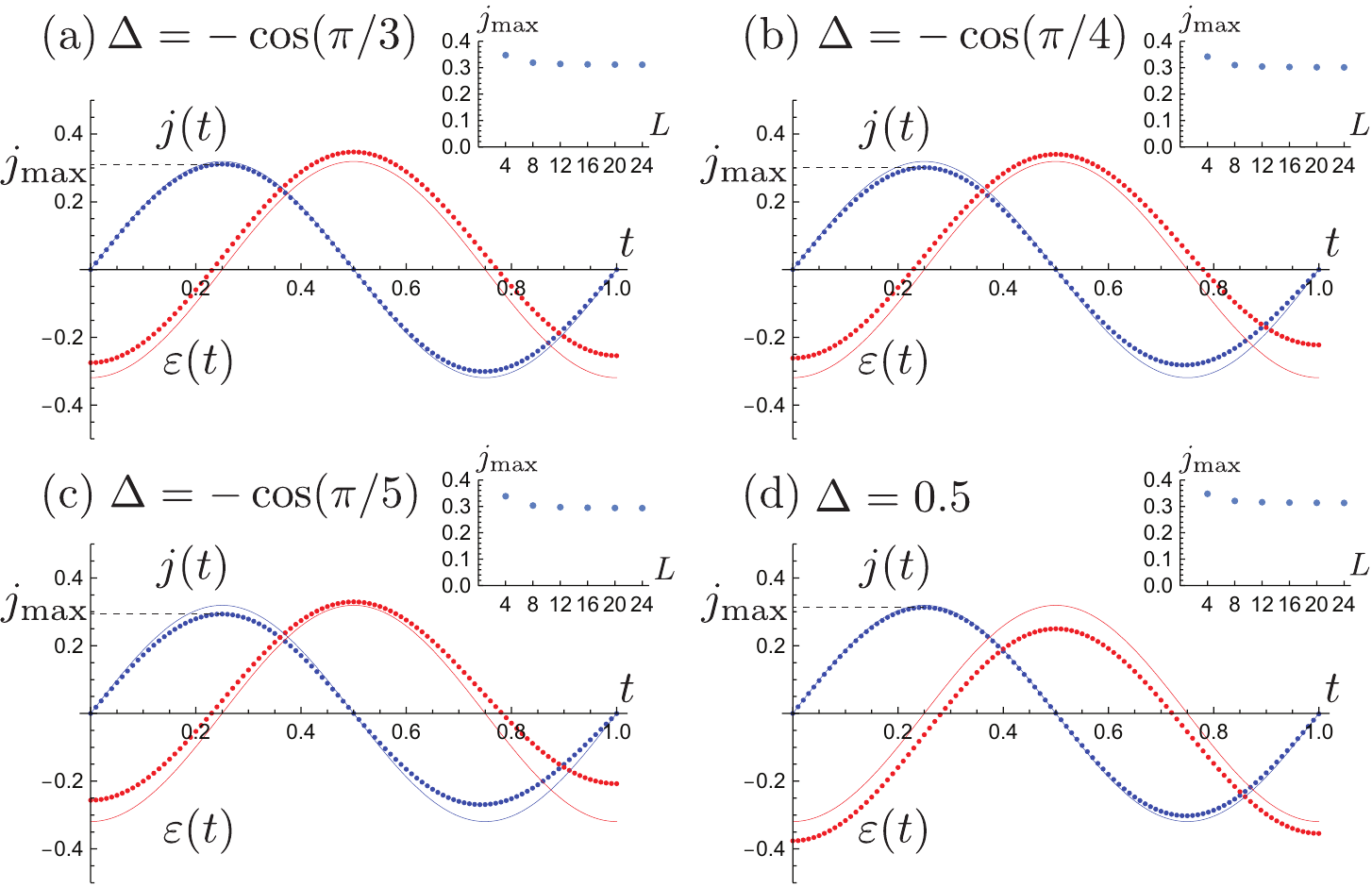}
\caption{Same as Fig.~\ref{fig4} but for $T=1$, which corresponds to a strong electric field $E=2\pi$ in the unit of $e/a$. 
Solid curves represent the analytic expressions for $\Delta=0$ for comparison (see the caption of Fig~\ref{fig1}).}
\label{fig5}
\end{center}
\end{figure}

\subsection{Bloch oscillations under a stronger field}
The subtle differences caused by $\Delta$ discussed above are irrelevant under a strong electric field. Indeed, we find qualitatively the same behavior regardless of the value of $\Delta$ for the $T=1$ case as shown in Fig.~\ref{fig5}.  The corresponding electric field $E=2\pi$ is so strong that non-adiabatic transitions occur everywhere. 

We also present the behavior of $\varepsilon(t)$ and $j(t)$ for $T=1, 10, 10^2,\cdots, 10^6$ for comparison in Fig.~\ref{fig7} in Appendix.
In principle, one can analyze the Landau-Zener transition probability for each level repulsion point, but these plots imply that behaviors of the $\varepsilon(t)$ and $j(t)$ depend on the details of the system except for the two limiting cases of adiabatic and non-adiabatic transitions.

\section{Discussions}
In this work, we demonstrated that interacting electrons can in principle exhibit Bloch oscillations even under a weak electric field when the interaction is fine-tuned. 
This is the case when the non-linear Drude weights introduced in Ref.~\onlinecite{PhysRevB.102.165137} remain finite in the large $L$ limit to all orders of response.
The expected behavior of the oscillations in the electric current density is similar to the noninteracting limit in the sense that the amplitude of the oscillation remains nonzero in the limit of large system size. 
Yet, there are also some differences, for example, in the number of peaks and the period of the oscillations for a given adiabatic process, as one can see by comparing Fig.~\ref{fig1} (b) and Fig.~\ref{fig4} (a)--(c).  In particular, the frequency of the oscillation becomes $a(p-1)E/(2\pi)$ for $\Delta=-\cos(\pi/p)$.

When the interaction is not exactly at the fine-tuned value, the oscillation in the limit of the weak electric field is suppressed in a large but finite system [Fig.~\ref{fig4} (d)]. 
However, even in that case, when the applied electric field is not weak enough, non-adiabatic transitions occur and one observes nontrivial oscillations of the electric current density, as shown in Fig.~\ref{fig5}.  In fact, our estimate based on the Landau-Zener formula in \eqref{LZ2} suggests that the adiabatic limit is difficult to be realized in a thermodynamically large system. This, in turn, implies that pathological behaviors of nonlinear Drude weights observed in Refs.~\cite{PhysRevB.102.165137,PhysRevB.103.L201120,Fava} do not come into play for a realistic strength of the applied electric field.

\begin{acknowledgments}
We thank Kazuaki Takasan and Masaki Oshikawa for collaborations in previous related works. 
We thank Hosho Katsura for partially informing us of their on-going result based on the Bethe ansatz, which supports the finiteness of nonlinear Drude weights we numerically observed~\cite{PC}.
We also thank Masaki Oshikawa and Hosho Katsura for their valuable comments on the earlier version of the draft.
The work of Y. F. is supported by JSPS KAKENHI Grant No.~JP20K14402.
The work of H. W. is supported by JSPS KAKENHI Grant No.~JP20H01825 and by JST PRESTO Grant No.~JPMJPR18LA. 
\end{acknowledgments}

\bibliography{ref.bib}

\appendix

\clearpage

\onecolumngrid

\section{Analytic formulas}
\label{sec:app}
Here we summarize the analytic expressions derived in previous studies. We parametrize $\Delta$ as $\Delta=\cos\gamma$ with $0\leq\gamma<\pi$.
The ground state energy density $\bar{\varepsilon}_{\text{GS}}$ and the linear Drude weight $\bar{\mathcal{D}}_1$ in the large $L$ limit are given by ~\cite{Hamer_1987,PhysRevLett.65.1833}
\begin{align}
\bar{\varepsilon}_{\text{GS}} &=\frac{1}{4}\cos\gamma-\frac{1}{2}\sin\gamma\int_{-\infty}^\infty dx\frac{\sinh[(\pi-\gamma) x]}{\sinh(\pi x)\cosh(\gamma x)},\\
\bar{\mathcal{D}}_1&=\frac{Kv}{\pi}=\frac{\pi}{4}\frac{\sin\gamma}{\gamma(\pi-\gamma)}.
\end{align}
The Luttinger parameter $K$ and the velocity parameter $v$ are given by~\cite{doi:10.1142/S0217979212440092,Giamarchi2004B}
\begin{align}
K&=\frac{\pi}{2(\pi-\gamma)},\\
v&=\frac{\pi\sin{\gamma}}{2\gamma}.
\end{align}
Higher-oder Drude weights $\bar{\mathcal{D}}_3$~\cite{PhysRevB.102.165137} and $\bar{\mathcal{D}}_5$~\cite{PhysRevB.102.165137,PhysRevB.103.L201120} are given by
\begin{align}
\bar{\mathcal{D}}_3&=-\frac{\sin\gamma}{16\gamma(\pi-\gamma)}\biggl(
\frac{
\Gamma\big(\frac{3\pi}{2\gamma}\big)\Gamma\big(\frac{\pi-\gamma}{2\gamma}\big)^3
}{
\Gamma\big(\frac{3(\pi-\gamma)}{2\gamma}\big)\Gamma\big(\frac{\pi}{2\gamma}\big)^3
}+\frac{3\pi\tan\big(\frac{\pi^2}{2\gamma}\big)}{\pi-\gamma}\biggr),\\
\bar{\mathcal{D}}_5=&\frac{3\sin{\gamma}}{64\pi\gamma(\pi-\gamma)}
\biggl(
\frac{\Gamma\big(\frac{5\pi}{2\gamma}\big){\Gamma\big(\frac{\pi-\gamma}{2\gamma}\big)}^5}{\Gamma\big(\frac{5(\pi-\gamma)}{2\gamma}\big){\Gamma\big(\frac{\pi}{2\gamma}\big)}^5}
-\frac{5\Gamma\big(\frac{3\pi}{2\gamma}\big)^2{\Gamma\big(\frac{\pi-\gamma}{2\gamma}\big)}^6}{3\Gamma\big(\frac{3(\pi-\gamma)}{2\gamma}\big)^2{\Gamma\big(\frac{\pi}{2\gamma}\big)}^6}+\frac{15\pi^2\tan^2{\big(\frac{\pi^{2}}{2\gamma}\big)}}{(\pi-\gamma)^2}
+\frac{5\pi\tan{\big(\frac{\pi^{2}}{2\gamma}\big)}}{\pi-\gamma}\frac{\Gamma\big(\frac{3\pi}{2\gamma}\big){\Gamma\big(\frac{\pi-\gamma}{2\gamma}\big)}^3}{\Gamma\big(\frac{3(\pi-\gamma)}{2\gamma}\big){\Gamma\big(\frac{\pi}{2\gamma}\big)}^3}
\biggr).
\end{align}
The formulas for $\bar{\mathcal{D}}_3$ and $\bar{\mathcal{D}}_5$ are finite when $\pi/3<\gamma<\pi$ and $\pi/2<\gamma<\pi$, respectively.

The coefficient $\lambda$ of  the least irrelevant perturbation due to Umklapp scattering derived in Ref.~\onlinecite{LUKYANOV1998533} reads
\begin{align}
\lambda=\frac{4\Gamma(\beta^{-2})}{\Gamma(1-\beta^{-2})}\biggl(\frac{\Gamma(1+\frac{\beta^2}{2-2\beta^2})}{2\sqrt{\pi}\Gamma(1+\frac{1}{2-2\beta^2})}\biggr)^{2/\beta^2-2}=\frac{4\Gamma(\frac{\pi}{\pi-\gamma})}{\Gamma(\frac{-\gamma}{\pi-\gamma})}\biggl(\frac{\Gamma(\frac{\pi+\gamma}{2\gamma})}{2\sqrt{\pi}\Gamma(\frac{\pi+2\gamma}{2\gamma})}\biggr)^{4K-2},
\end{align}
where $\beta^2=1-\gamma/\pi$, which gives
\begin{align}
C=|\lambda|\frac{\pi\sin\gamma}{2\gamma}(2\pi)^{4K-1}=4\pi^2\frac{\sin\gamma}{\gamma}\frac{\Gamma(\frac{\pi}{\pi-\gamma})}{\big|\Gamma(\frac{-\gamma}{\pi-\gamma})\big|}\biggl(\sqrt{\pi}\frac{\Gamma(\frac{\pi+\gamma}{2\gamma})}{\Gamma(\frac{\pi+2\gamma}{2\gamma})}\biggr)^{4K-2}.
\end{align}

\begin{figure}[t]
\begin{center}
\includegraphics[width=0.99\columnwidth]{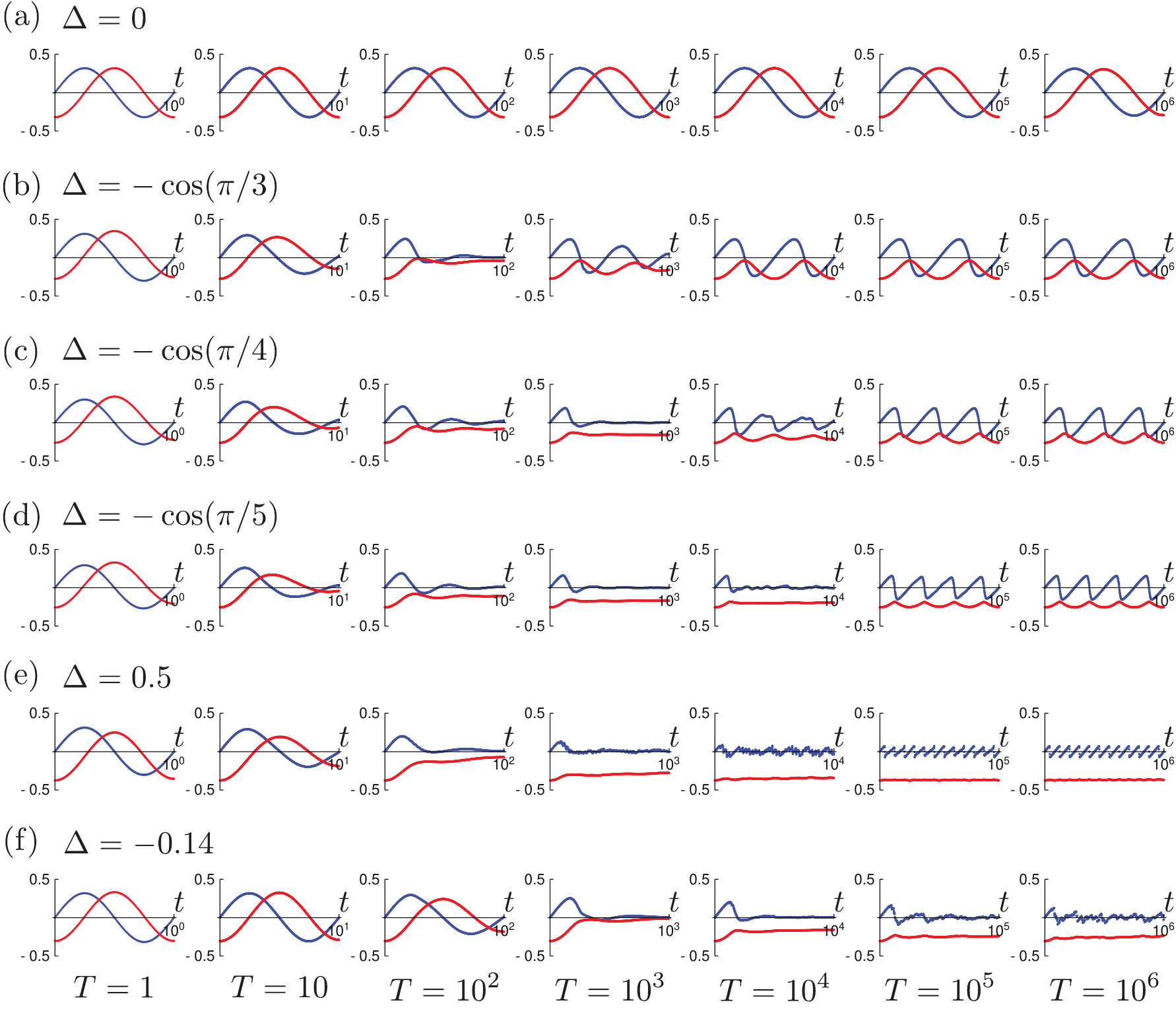}
\caption{
The plot of the energy density $\varepsilon(t)$ (red) and the current density $j(t)$ (blue) under the electric field $E=2\pi/T$ for various values of $T$ and $\Delta$:
$T=1$, $10$, $10^2$, $\cdots$, and $10^6$ and $\Delta=0$, $-\cos(\pi/3)$, $-\cos(\pi/4)$, $-\cos(\pi/5)$, $0.5$, and $-0.14$. 
The system size is $L=24$. While the gap above the adiabatic ground state is maximized at $\Delta=-0.14$ for $\Delta<0$ and $L=24$,  $T=10^6$ is not large enough to see $O(L)$ oscillations as observed for $\Delta=0.5$.  }
\label{fig7}
\end{center}
\end{figure}

\clearpage

\end{document}